\documentclass[aps,prd,preprint,superscriptaddress,amsmath,amssymb,showpacs]{revtex4-1}
\usepackage{dcolumn}
\usepackage{graphicx}
\usepackage{float}
\usepackage{physics}
\usepackage[colorlinks=true,allcolors=blue]{hyperref}
\usepackage{amsmath,graphicx,amsfonts, mathrsfs,amssymb}
\usepackage{amsmath,epsfig}
\usepackage{amssymb,amsfonts}
\usepackage{color}
\usepackage{latexsym}
\usepackage{epsfig}

\begin{document}

\title{Holographic Schwinger Effect in Anisotropic Media}
\author{Jing Zhou}
\affiliation{Department of Physics, Hunan City University, Yiyang, Hunan 413000, China}
\affiliation{All-solid-state Energy Storage Materials and Devices Key Laboratory of Hunan Province, Hunan City University, Yiyang 413000, China}
\affiliation{Department of Physics, Nanjing Normal University, Nanjing, Jiangsu 210097, China}
\author{Jun Chen}
\affiliation{Department of Physics, Hubei Minzu University, Enshi 445000, China}
\author{Le Zhang}
\affiliation{College of Physics and Electronic Science, Hubei Normal University, Huangshi 435002, China}
\author{Jialun Ping}
\affiliation{Department of Physics, Nanjing Normal University, Nanjing, Jiangsu 210097, China}
\author{Xun Chen}
\email{chenxunhep@qq.com}
\affiliation{School of Nuclear Science and Technology, University of South China, Hengyang 421001, China}
\begin{abstract}
According to gauge/gravity correspondence, we study the holographic Schwinger effect within an anisotropic background. Firstly, the separate length of the particle-antiparticle pairs is computed within the context of an anisotropic background which is parameterized by dynamical exponent $\nu$. It is found that the maximum separate length $x$ increases with the increase of dynamical exponent $\nu$.  By analyzing the potential energy, we find that the potential barrier increases with the dynamical exponent $\nu$ at a small separate distance. This observation implies that the Schwinger effect within an anisotropic background is comparatively weaker when contrasted with its manifestation in an isotropic background.
Finally, we also find that the Schwinger effect in the transverse direction is weakened compared to the parallel direction in the anisotropic background, which is consistent with the top-down model.

\end{abstract}
\maketitle

\section{Introduction}\label{sec:01_intro}
\label{intro}
\setlength{\parskip}{1em}
It is known that the pair production of electron and positron under a strong external electric field is named as Schwinger effect\cite{Schwinger:1951nm}. This phenomenon shows a general feature of vacuum instability in the presence of the external field. A qualitative understanding of this phenomenon can be obtained by looking at the potential energy of the pair in the presence of an electric field $E$\cite{Sato:2013iua}
\begin{eqnarray}
V(x) & = & 2 m-E x-\frac{\alpha_{\mathrm{s}}}{x},
\end{eqnarray}
where $\alpha_{s} \simeq 1 / 137$ is the fine-structure constant and virtual pairs are separated by a distance x. The pair production is described as a tunneling process that creates a particle-antiparticle pair. From the above formula, one can find that the potential barrier decreases with the increase of electric field, and vanishes at a certain critical electric field $E_{c}$. It is instructive to mention that,
comparing with the Schwinger case, one can find that there exists a critical field $e E_{c} = \left(4 \pi / e^{2}\right) m^{2}$  in the Affleck-Alvarez-Manton(AAM)\cite{Affleck:1981bma} case. The critical value does not satisfy the weak-field condition $e E \ll m^{2}$.

Anti-de Sitter/Conformal Field Theory correspondence\cite{Maldacena:1997re,Witten:1998qj,Aharony:1999ti,Witten:1998zw}provides a way to study the Schwinger effect at the strong coupling and there is no constraint for the values of external fields\cite{Gorsky:2001up,Semenoff:2011ng}. Then it is natural to consider the Schwinger effect in the holographic method. The holographic Schwinger effect was formally proposed in Ref.\cite{Semenoff:2011ng} and they studied the particles produced in the $\mathcal{N}$=4 super-Yang Mills theory which is dual to the N D3-brane with a probe D3-brane placed at a finite radial position in the bulk \cite{Wu:2015krf}. In the usual studies, the test particles
are assumed to be a heavy quark limit. To avoid pair creation suppressed by the divergent mass, the location of the probe D3-brane is at a finite radial position rather than at the AdS boundary\cite{Semenoff:2011ng,Zhu:2019igg}.
Following this idea, lots of works have been carried out to study the holographic Schwinger effect\cite{Sato:2013pxa,Zhang:2020noe,Zhang:2017egb,Cai:2016jgr,Sonner:2013mba,Ambjorn:2011wz,Zhang:2015bha,Ghoroku:2016kft,Zhang:2018oie,Dehghani:2015gtd,Sato:2013hyw,Sato:2013dwa,Kawai:2013xya,Kawai:2015mha}.

In Ref.\cite{Zhu:2019igg,Zhang:2018hfd}, they discuss the holographic Schwinger effect in the extreme conditions created in the high-energy physics experiment. In addition to high temperature, large chemical potential, and strong magnetic field, partonic system generated in ultra-relativistic heavy-ion collisions can not be homogeneous and isotropic at the very early time of collision\cite{Strickland:2013uga}. Asymptotic weak-coupling enhances the longitudinal expansion substantially more than the radial expansion so the system becomes colder in the longitudinal direction than in the transverse direction\cite{Thakur:2012eb}.  Thus, the momenta of the partons along the longitudinal direction are lower than that in the transverse direction. In other words, in the anisotropic stage, the longitudinal and transverse pressures satisfy $P_{L}<P_{T}$ with the corresponding momenta of the partons $\left\langle p_{L}^{2}\right\rangle<\left\langle p_{T}^{2}\right\rangle$. Moreover,
the experimental energy dependence of the total multiplicity can be reproduced in the anisotropic background, but all attempts to reproduce this dependence in isotropic models failed \cite{Arefeva:2018hyo}.
Therefore, it is inspired to know how the holographic Schwinger effect performs in an anisotropic background.

Some interesting results of anisotropy through the holographic method have been carried out in recent years. For example, the thermodynamics and instabilities of anisotropic plasma are discussed in Ref.\cite{Mateos:2011tv}. In Refs.\cite{Chernicoff:2012iq,Chernicoff:2012gu}, they study the jet quenching and drag force in anisotropic plasma. Thermal photon production in the anisotropic plasma was also researched in Ref.\cite{Patino:2012py}. The quarkonium dissociation was discussed in Ref.\cite{Chernicoff:2012bu}. In particular, the anisotropic background discloses a more abundant structure than that in the isotropic case with the small/large black holes phase transition\cite{Arefeva:2018cli}. Other related works can be found in Refs.\cite{Giataganas:2012zy,Chakrabortty:2013kra,Ali-Akbari:2013txa,Misobuchi:2015ioa,Itsios:2018hff,Jahnke:2017iwi,Chakraborty:2017msh,Finazzo:2016mhm,Arefeva:1993rz,deForcrand:2017fky}.

In this work, we mainly focus on the holographic Schwinger effect in the anisotropic 5-dimensional Einstein-dilaton-two-Maxwell system \cite{Arefeva:2018hyo} and the anisotropic background is parameterized by dynamical exponent $\nu$. The remainder of this paper is organized as follows: In Sec.\ref{sec:02_Background geometry}, we introduce the 5-dimensional Einstein-dilaton-two-Maxwell system. In Sec.\ref{sec:03}, we mainly focus on the potential analysis in anisotropic background. In Sec.\ref{sec:04}, we study the potential analysis in finite chemical potential and different warp factor coefficients. The conclusion can be found in Sec.\ref{sec:06}.
\section{Background geometry}\label{sec:02_Background geometry}
The 5-dimensional Einstein-dilaton-two-Maxwell system was introduced in Ref.\cite{Arefeva:2018hyo}, which describes an anisotropic background parameterized by the dynamical exponent $\nu$. This background can give the total multiplicity dependence on energy, which agrees with the experimental data \cite{Arefeva:2018hyo}. And the action in the Einstein
frame is given by
\begin{equation}
S=\int \frac{d^{5} x}{16 \pi G_{5}} \sqrt{-\operatorname{det}\left(g_{\mu \nu}\right)}\left[R-\frac{f_{1}(\phi)}{4} F_{(1)}^{2}-\frac{f_{2}(\phi)}{4} F_{(2)}^{2}-\frac{1}{2} \partial_{\mu} \phi \partial^{\mu} \phi-V(\phi)\right],
\end{equation}
where $F_{1}$ is Maxwell field with field strength tensor $F_{\mu \nu}^{(1)}  =  \partial_{\mu} A_{\nu}-\partial_{\nu} A_{\mu}$, and $F_{2}$ is the other Maxwell field with field strength tensor $F_{\mu \nu}^{(2)}=q d y^{1} \wedge d y^{2}$. $f_{1}(\phi)$, $f_{2}(\phi)$ are the gauge functions that correspond to the two Maxwell fields. $V(\phi)$ is the scalar potential. The metric ansatz of the black brane solution in the anisotropic background is
\begin{equation}
\begin{aligned}
d s^{2}=\frac{L^{2} b(z)}{z^{2}}\left[-g(z) d t^{2}+d x^{2}+z^{2-\frac{2}{\nu}}\left(d y_{1}^{2}+d y_{2}^{2}\right)+\frac{d z^{2}}{g(z)}\right] \\ \phi=\phi(z), \quad A_{\mu}^{(1)}=A_{t}(z) \delta_{\mu}^{0} \\ F_{\mu \nu}^{(2)}=q d y^{1} \wedge d y^{2},
\end{aligned}
\end{equation}
where $b(z)=e^{cz^2/2}$ is the warp factor, and $c$ represents the deviation from conformality. g(z) is the blackening function. As we know, the hot matter produced in the early stage of relativistic heavy ion collisions is anisotropic where the longitudinal and transverse expansion are different. Therefore, our goal is to use the anisotropic metric to qualitatively simulate this anisotropy of relativistic heavy ion collisions by the dynamical exponent $\nu$.
Following Ref.\cite{Arefeva:2018hyo}, we take all physical quantities as dimensionless units and set the AdS radius $L$ to one. By solving the equation of motion obtained from the above action, the function $g(z)$ can be calculated as
\begin{equation}
g(z)=1-\frac{z^{2+\frac{2}{\nu}}}{z_{h}^{2+\frac{2}{\nu}}} \frac{\mathfrak{G}\left(\frac{3}{4} c z^{2}\right)}{\mathfrak{G}\left(\frac{3}{4} c z_{h}^{2}\right)}-\frac{\mu^{2} c z^{2+\frac{2}{\nu}} e^{\frac{c z_{h}^{2}}{2}}}{4\left(1-e^{\frac{c z_{h}^{2}}{4}}\right)^{2}} \mathfrak{G}\left(c z^{2}\right)+\frac{\mu^{2} c z^{2+\frac{2}{\nu}} e^{\frac{c z_{h}^{2}}{2}}}{4\left(1-e^{\frac{c z_{h}^{2}}{4}}\right)^{2}} \frac{\mathfrak{G}\left(\frac{3}{4} c z^{2}\right)}{\mathfrak{G}\left(\frac{3}{4} c z_{h}^{2}\right)} \mathfrak{G}\left(c z_{h}^{2}\right),
\end{equation}
and
\begin{equation}
\mathfrak{G}(x)=\sum_{n=0}^{\infty} \frac{(-1)^{n} x^{n}}{n !\left(1+n+\frac{1}{\nu}\right)}.
\end{equation}
Then the temperature can be given as
\begin{equation}
T\left(z_{h}, \mu, c, \nu\right)=\frac{g^{\prime}\left(z_{h}\right)}{4 \pi}=\frac{e^{-\frac{3 c z_{h}^{2}}{4}}}{2 \pi z_{h}}\left|\frac{1}{\mathfrak{G}\left(\frac{3}{4} c z_{h}^{2}\right)}+\frac{\mu^{2} c z_{h}^{2+\frac{2}{\nu}} e^{\frac{c z_{h}}{4}}}{4\left(1-e^{\frac{c z_{h}^{2}}{4}}\right)^{2}}\left(1-e^{\frac{c z_{h}^{2}}{4}} \frac{\mathfrak{G}\left(c z_{h}^{2}\right)}{\mathfrak{G}\left(\frac{3}{4} c z_{h}^{2}\right)}\right)\right|.
\end{equation}
\section{Potential analysis in anisotropic background}\label{sec:03}
The coordinates of the particle pairs in parallel case can be written as
\begin{eqnarray}
t= \tau, \quad y_{1}=\sigma, \quad x= y_{2}= 0, \quad z =z(\sigma).
\end{eqnarray}
Similarly, the transverse direction is defined as
\begin{eqnarray}
t= \tau, \quad x=\sigma, \quad y_{1}= y_{2}= 0, \quad z =z(\sigma).
\end{eqnarray}
The Nambu-Goto action reads
\begin{eqnarray}
S= T_{F} \int d \sigma d \tau \mathcal{L}= T_{F} \int d \sigma d \tau \sqrt{-\operatorname{det} g_{\alpha \beta}},
\end{eqnarray}
where $T_{F}$ is the string tension and $g_{\alpha \beta}$ is the induced metric. The Lagrangian density in the parallel case can be written as
\begin{eqnarray}
\mathcal{L}=\sqrt{\operatorname{det} g_{\alpha \beta}}= \frac{b(z)}{z^2}\sqrt{g(z)z^{2-\frac{2}{\nu}}+\dot{z}^2}.
\end{eqnarray}
$\mathcal{L}$ does not rely on $\sigma$, so it must satisfy the follow equation
\begin{eqnarray}
\mathcal{L}-\frac{\partial \mathcal{L}}{\partial \dot{z}} \dot{z}=C. \label{conserved quantity}
\end{eqnarray}
Moreover, the boundary condition gives
\begin{eqnarray}
\frac{d z}{d \sigma}= 0, \quad z= z_{c}\left(z_{h}<z_{c}<z_{0}\right),
\end{eqnarray}
here one should note that the probe D3 brane locates at $z=z_{0}$. So, the conserved quantity can be evaluated as
\begin{eqnarray}
C =\frac{b(z_{c})}{z_{c}^{2}}\sqrt {g(z_{c})z_{c}^{2-\frac{2}{\nu}} }. \label{c}
\end{eqnarray}
Combining Eq.~\ref{conserved quantity} and Eq.~\ref{c}, one finds
\begin{eqnarray}
\dot{z} =\frac{d z}{d \sigma}={\sqrt{g(z)z^{2-\frac{2}{\nu}}\left(\frac{\frac{b(z)^2g(z)z^{2-\frac{2}{\nu}}}{z^4}}{\frac{b(z_{c})^2 g(z_{c})z_{c}^{2-\frac{2}{\nu}}}{z_{c}^4}}-1\right)}}, \label{zz}
\end{eqnarray}
then the separating length of the test particle pairs can be obtained by integrating Eq.~\ref{zz}
\begin{eqnarray}
x= 2 \int_{z_{c}}^{z_{0}} dz \frac{1}{\sqrt{g(z)z^{2-\frac{2}{\nu}}\left(\frac{\frac{b(z)^2g(z)z^{2-\frac{2}{\nu}}}{z^4}}{\frac{b(z_{c})^2 g(z_{c})z_{c}^{2-\frac{2}{\nu}}}{z_{c}^4}}-1\right)}}. ~\label{separating}
\end{eqnarray}
With the separating length and Lagrangian density in hand, the sum of Coulomb potential and static energy is
\begin{equation}
\begin{aligned}
V_{(CP+S E)}&=2T_{F} \int_{0}^{\frac{x}{2}} d \sigma \mathcal{L} \\
&=2 T_{F}\int_{z_{c}}^{z_{0}} d z \frac{\sqrt{\frac{b(z)^2g(z)z^{2-\frac{2}{\nu}}}{z^4}\frac{b(z)^2}{z^4} }}{\sqrt{\frac{b(z)^2g(z)z^{2-\frac{2}{\nu}}}{z^4}-\frac{b(z_{c})^2 g(z_{c})z_{c}^{2-\frac{2}{\nu}}}{z_{c}^4}}}. ~\label{vtot}
\end{aligned}
\end{equation}
To obtain the critical electric field, one should compute the DBI action of the probe D3 brane, namely
\begin{equation}
S_{DBI}= -T_{D 3} \int d^{4} x \sqrt{-\operatorname{det}\left(G_{\mu \nu}+\mathcal{F}_{\mu \nu}\right)}.~\label{DBI}
\end{equation}
To simplify the analysis, assuming that the external electric field is oriented along the $x$ direction, one can find
\begin{equation}
G_{\mu v}+\mathcal{F}_{\mu v}=\left(\begin{array}{cccc}
-g(z)\frac{b(z)}{z^2} & 2 \pi \alpha^{\prime} E & 0 & 0 \\
-2 \pi \alpha^{\prime} E & \frac{b(z)}{z^2} & 0 & 0 \\
0 & 0 & \frac{b(z)}{z^2}z^{2-\frac{2}{\nu}} & 0 \\
0 & 0 & 0 & \frac{b(z)}{z^2}z^{2-\frac{2}{\nu}}.
\end{array}\right)
\end{equation}
Then we can rewrite Eq.~\ref{DBI} at $z=z_{0}$ as
\begin{eqnarray}
S_{D B I} & = & -T_{D 3} \int d^{4} x \sqrt{b\left(z_{0}\right)^{2} z_{0}^{-4-\frac{4}{\nu}}} \sqrt{-b\left(z_{0}\right)^{2} g\left(z_{0}\right)+\left(2 \pi \alpha^{\prime} z_{0}^{2}\right)^{2} E^{2}} .
\end{eqnarray}
If the equation has a physical meaning, then we require
\begin{eqnarray}
-b(z_{0})^2g(z_{0})+\left(2 \pi \alpha^{\prime}z_{0}^{2}\right)^{2} E^{2} \geq 0.
\end{eqnarray}
By simple calculation, one can find that the critical electric field is
\begin{eqnarray}
E_{c} =T_{F}\sqrt{\frac{b(z_{0})^2g(z_{0})}{z_{0}^4}}.
\end{eqnarray}
If we define a dimensionless parameter $\beta \equiv \frac{E}{E_{c}}$, then the total potential of the particle-antiparticle pair will be
\begin{equation}
\begin{aligned}
V_{t o t}=&V_{(C P+S E)}-Ex\\
         =&2 T_{F}\int_{z_{c}}^{z_{0}} d z \frac{\sqrt{\frac{b(z)^2g(z)z^{2-\frac{2}{\nu}}}{z^4}\frac{b(z)^2}{z^4} }}{\sqrt{\frac{b(z)^2g(z)z^{2-\frac{2}{\nu}}}{z^4}-\frac{b(z_{c})^2 g(z_{c})z_{c}^{2-\frac{2}{\nu}}}{z_{c}^4}}}
          -2 T_{F}\beta \int_{z_{c}}^{z_{0}} dz \frac{\sqrt{\frac{b(z_{0})^2g(z_{0})}{z_{0}^4}}}{\sqrt{g(z)z^{2-\frac{2}{\nu}}\left(\frac{\frac{b(z)^2g(z)z^{2-\frac{2}{\nu}}}{z^4}}{\frac{b(z_{c})^2 g(z_{c})z_{c}^{2-\frac{2}{\nu}}}{z_{c}^4}}-1\right)}}. ~\label{V}
\end{aligned}
\end{equation}
The transverse case can be calculated in a similar way. Now we can investigate the Schwinger effect in the anisotropic background. First of all, we calculate the separating length $x$ which is given by Eq.~\ref{separating}.
\begin{figure}
\centering
   \resizebox{0.6\textwidth}{!}{
    \includegraphics{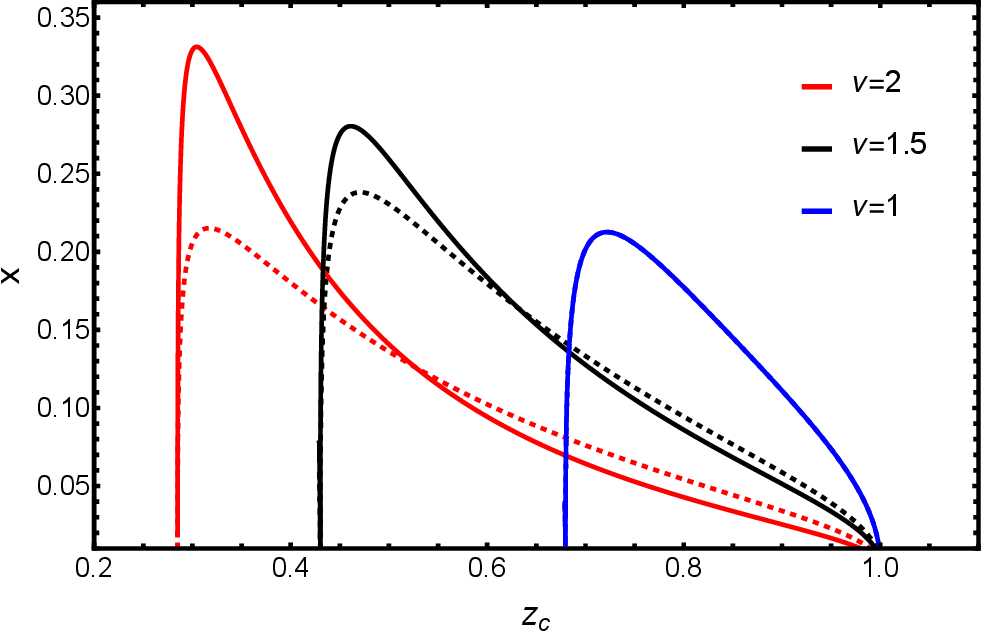}}
   \caption{\label{anisoXvsr}The separating length $x$ as a function of $z_{c}$ with the given values of the chemical potential $\mu=3$, the temperature $T =0.5$ and the warp factor coefficient $c=-0.3$. The transverse direction is indicated by the dashed line, while the parallel direction is denoted by the solid line. The blue line is $\nu= 1$, the black line is $\nu = 1.5$ and the red line is $\nu= 2$, respectively. }
\end{figure}
The dependence of separating length on $z_{c}$ is shown in Fig.~\ref{anisoXvsr}. For diverse values of the dynamical exponent $\nu$, it is evident that the U-shaped string configuration manifests instability at smaller $z_{c}$ values while displaying stable behavior at larger $z_{c}$ values. Note that space attains isotropy when $\nu= 1$. From the picture, we can find that the maximum value of separating length increases with the dynamical exponent $\nu$. Then it may indicate that the Schwinger effect is weakened in the anisotropic background compared with the isotropic one. Furthermore, our investigation demonstrated that in the absence of an external electric field (in the $x$-direction), the maximum value of separating length in the parallel direction exceeds that in the transverse direction.
\begin{figure}
\centering
   \resizebox{1\textwidth}{!}{
   \includegraphics{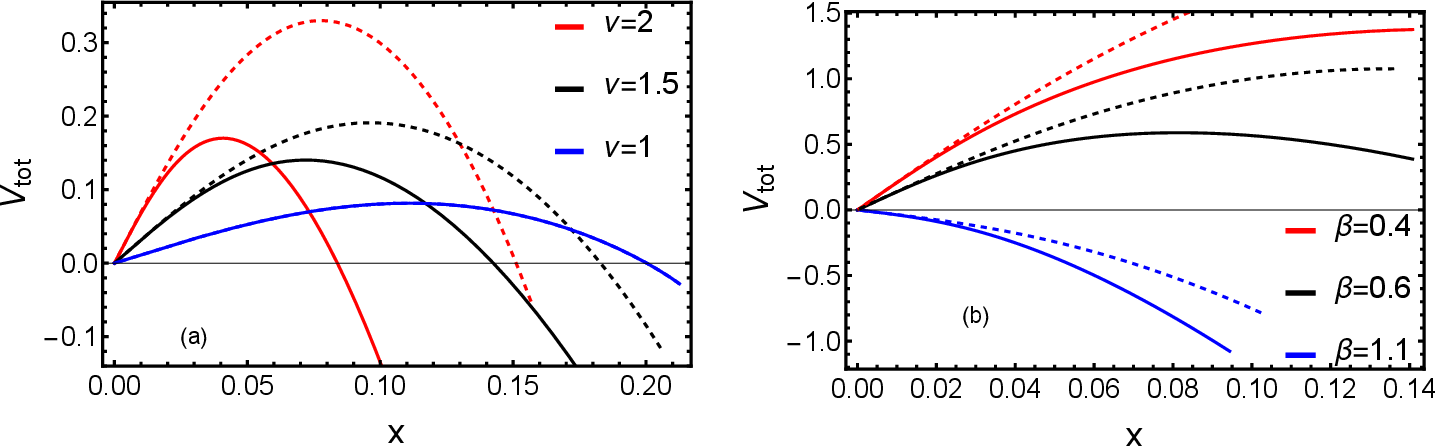}}
    \caption{\label{anisoVvsx}
    (a)The total potential $V_{tot}$ as a function of separating length $x$ for different dynamical exponent $\nu$ with the given values of the chemical potential $\mu=3$, the temperature $T = 0.5$ and the warp factor coefficient $c=-0.3$. The transverse direction is indicated by the dashed line, while the parallel direction is denoted by the solid line. The red line is $\nu=2$, the black line is $\nu=1.5$ and the blue line is $\nu=1$, respectively. (b) The total potential $V_{tot}$ against separating length x at different $\beta$ with $\nu=2$. The red line is $\beta=0.4$, black line is $\beta=0.6$ and blue line is $\beta=1.1.$ }
\end{figure}
\begin{figure}
\centering
   \resizebox{0.6\textwidth}{!}{
    \includegraphics{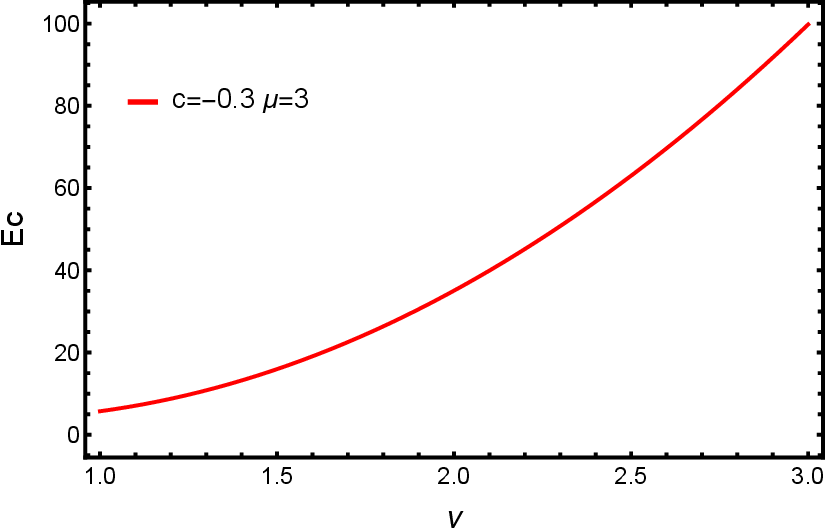}}
    \caption{\label{Ev} $E_{c}$ versus dynamical exponent $\nu$ with the given values of the temperature $T=0.5$, warp factor coefficient $c=-0.3$ and the chemical potential $\mu=3$.
    }
\end{figure}

Utilizing Eq.~\ref{vtot}, we analyze the correlation between the total potential of a particle-antiparticle pair and the separation length $x$, as shown in Fig~\ref{anisoVvsx}.
We find that the potential barrier is amplified by the dynamical exponent $\nu$ at small values of separating length. This enhanced potential barrier translates to a weaker Schwinger effect. The result is in partly qualitative agreement with the top-down holographic approach of the anisotropic Schwinger effect in \cite{Li:2022hka}.
Under the condition of an externally applied electric field in the $x$-direction, we find that the maximum total potential energy in the transverse direction is greater than that in the parallel direction. This implies that the Schwinger effect in the transverse direction is attenuated relative to the parallel direction.
Furthermore, the potential barrier decreases with the increase of the external electric field. The decrease in the potential barrier signifies an increased propensity for particle production.
It is easy to find that the potential barrier is present when $E<E_{c}$$\left(\beta<1\right)$. In this context, particle production can be regarded as a tunneling process. As the electric field gradually surpasses the critical threshold, the vacuum will become very unstable and the potential barrier will vanish \cite{Sato:2013iua}.
According to Eq.~\ref{V}, if the total potential energy increases, the critical electric field will be an increasing function of the dynamic exponent. This trend is vividly demonstrated in Fig.\ref{Ev}.
This implies that virtual particles require more energy from the outside to become real particles. This observation not only corroborates the results presented in Fig.\ref{anisoVvsx} but also aligns with the conclusions put forth in Ref.\cite{Li:2022hka}. Because the external electric field is identical in both the parallel and transverse directions, we exclusively performed calculations for the parallel direction to provide a representative depiction.

\section{potential analysis with chemical potential and warp factor coefficient}\label{sec:04}
\begin{figure}
\centering
    \resizebox{0.6\textwidth}{!}{
    \includegraphics{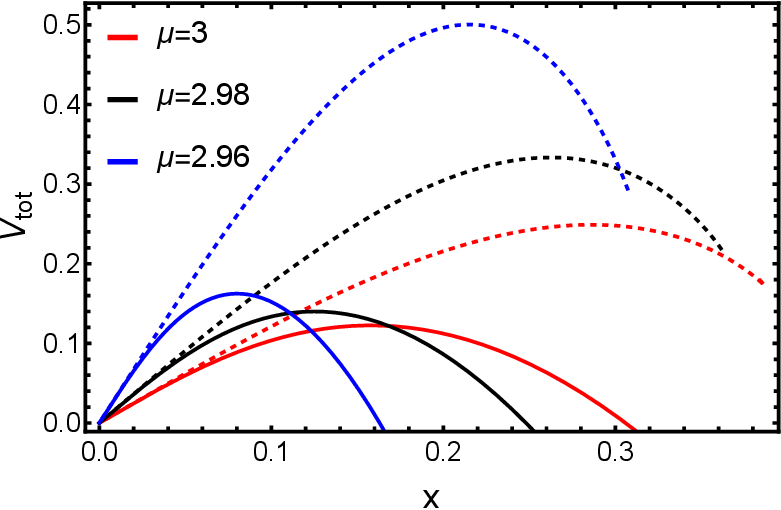}}
    \caption{\label{huaxueshibxvsL}
    The total potential $V_{tot}$ as a function of separating length x at difference chemical potential with the given values of the temperature $T = 0.58$, $\beta=0.8$, and the warp factor $c=-0.3$. The transverse direction is indicated by the dashed line, while the parallel direction is denoted by the solid line. The red line is $\mu=3$, the black line is $\mu=2.98$ and the blue line is $\mu=2.96$, respectively.
    }
\end{figure}
\begin{figure}
\centering
    \resizebox{0.6\textwidth}{!}{
    \includegraphics{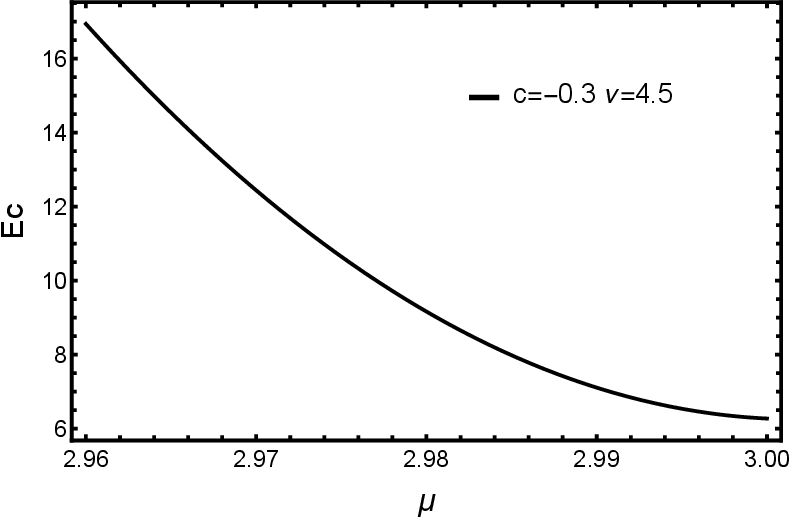}}
    \caption{\label{Eu} $E_{c}$ versus chemical potential $\mu$ with the given values of the temperature $T=0.58$, warp factor coefficient $c=-0.3$, and the dynamical exponent $\nu=4.5$.
    }
\end{figure}
The effect of the chemical potential on the total potential in different external electric fields is examined in Fig.~\ref{huaxueshibxvsL}. It is found that the total potential is reduced by the chemical potential in small distance $x$, particularly with $\beta=0.8$. Thus we we infer that the yield of particles increases with the increase of chemical potential, which is qualitatively consistent with the results in Ref.\cite{Zhang:2018hfd}.
One potential explanation for the enhancement of the Schwinger effect by a chemical potential is that the presence of the chemical potential provides an additional energy source for the created particles. This allows the particles to extract more energy from an external electric field, leading to an increase in the overall energy available for pair production and consequently enhancing the Schwinger effect.
Specifically, our results reveal a notable disparity between the total potential energy in the transverse and parallel directions. To be precise, the total potential energy in the transverse direction was found to be considerably higher compared to the parallel direction. As a result, this discrepancy indicates a correspondingly lower particle yield in the transverse direction.
Additionally, we also can find that the external critical field demonstrates a decrementing trend in response to the chemical potential in Fig.~\ref{Eu}.
This suggests that real particles are easier to produce in the presence of the external field which is consistent with the result in Fig.~\ref{huaxueshibxvsL}.

\begin{figure}
\centering
    \resizebox{0.6\textwidth}{!}{
    \includegraphics{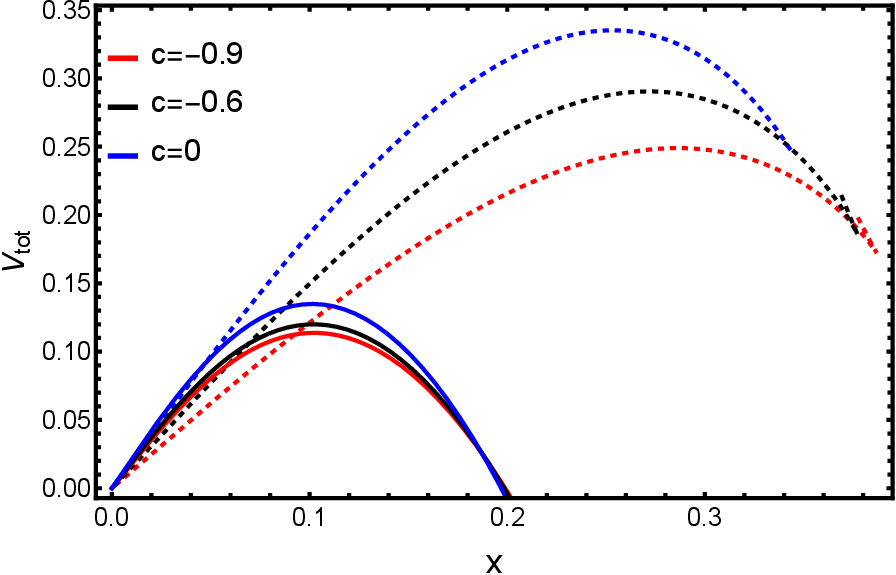}}
    \caption{\label{cVvsX}
    The total potential $V_{tot}$ versus separate length $x$ for different warp factor coefficients $c$. The temperature $T = 0.5$, $\beta=0.8$ and chemical potential $\mu=2$. The transverse direction is indicated by the dashed line, while the parallel direction is denoted by the solid line. The red line is $c=-0.9$, the black line is $c=-0.6$ and the blue line is $c=0$, respectively.
    }
\end{figure}
\begin{figure}
\centering
    \resizebox{0.6\textwidth}{!}{
    \includegraphics[width=8cm]{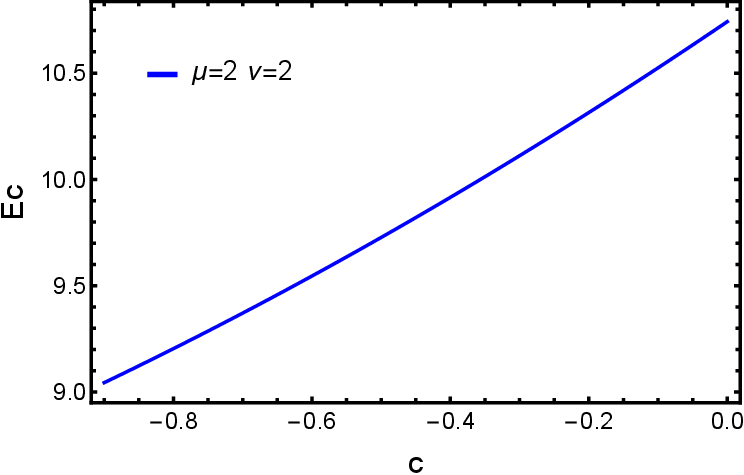}}
    \caption{\label{Ecc} $E_{c}$ versus warp factor coefficient $c$. The chemical potential $\mu=2$, the temperature $T = 0.5$ and the dynamical exponent $\nu=2$.
    }
\end{figure}

The effect of the warp factor coefficient on the total potential in different external electric fields is plotted in Fig.~\ref{cVvsX}. Here, $c$ signifies the degree of deviation from conformality. Notably, the warp factor coefficient is zero in the context of pure AdS, a conformal theory. However, the real QCD is inherently non-conformal, leading to non-zero values for $c$. Here we take $c=-0.6,-0.9$. In Fig.~\ref{cVvsX}, one can find that the total potential is reduced by warp factor coefficient $c$ in small distance $x$. This implies that the warp factor coefficient can reduce the Schwinger effect. Similar to the case with different chemical potentials, we find that the total potential energy in this scenario is also higher in the transverse direction compared to the parallel direction. To further explore this relationship, we depict $E_{c}$ as a function of $c$ in Fig.~\ref{Ecc}. The critical electric field increases with the increase of warp factor coefficient $c$. This implies that more energy needs to be obtained from the external electric field to overcome the potential barrier.

\section{Summary and Conclusions}\label{sec:06}
In this paper, we study the Schwinger effect in the Einstein-dilaton-two-Maxwell-scalar system in an anisotropic background. The anisotropic models can reappear properties such as the anisotropic pressure of QGP in heavy-ion collision. Then it is natural to study how the Schwinger effect is changed in the anisotropic case.

The separate length of the particle-antiparticle pair in the anisotropic background is computed. As the dynamical exponent $\nu$ rises, the U-shaped string exhibits instability at small $z_{c}$, while stability prevails at large $z_{c}$ values.
Through the utilization of the Dirac-Born-Infeld (DBI) action for probing D3 branes, we determine the critical electric field $E_{c}$ and compute the total potential. It is found that the dynamical exponent $\nu$ enlarges the potential barrier. This means that
the production of particles is suppressed. In comparison to the parallel direction, our findings indicate a reduction in the strength of the Schwinger effect in the transverse direction. We also find the critical electric field is reduced by the chemical potential but enhanced by the warp factor coefficient $c$ and the dynamical exponent.

Since the Schwinger effect is an important mechanism to create a plasma of gluons and quarks from
initial color-electric flux tubes \cite{Oliva:2016ltp}, we hope that the Schwinger effect in the anisotropic background could provide some new insights into the understanding of the QGP. Moreover, the potential analysis in the holographic shock wave model may be worth discussing~\cite{Arefeva:2015jkr} in future work.
\section*{Acknowledgments}
This work is partly supported by the National Natural Science Foundation of China under Contract Nos. 11775118, 11535005,12005056,11947050, the Natural Science Foundation of Hunan Province of China under Grant
No.2022JJ40344 and the Research Foundation of Education Bureau of Hunan Province, China under Grant Nos. 21B0402 and 22B0788.

\end{document}